\documentclass[conference]{IEEEtran}
\IEEEoverridecommandlockouts
\usepackage{cite}
\usepackage{amsmath,amssymb,amsfonts}
\usepackage{algorithmic}
\usepackage{graphicx}
\usepackage{textcomp}
\usepackage{xcolor}
\usepackage{hyperref}
\usepackage{tabularx}
\usepackage[most, theorems]{tcolorbox}
\usepackage{multirow}
\usepackage{booktabs}

\def\BibTeX{{\rm B\kern-.05em{\sc i\kern-.025em b}\kern-.08em
    T\kern-.1667em\lower.7ex\hbox{E}\kern-.125emX}}

\usepackage{listings}
\usepackage{booktabs}
\usepackage{microtype}

\usepackage{multirow}     
\usepackage{colortbl}     
\newcommand{\imp}[1]{\textcolor{green!50!black}{$\checkmark$}\,$+#1$}
\newcommand{\wor}[1]{\textcolor{red!80!black}{$\times$}\,$-#1$}

\newcommand{\upG}{\hspace{0.22em}\textcolor{green!50!black}{$\uparrow$}}
\newcommand{\dnG}{\hspace{0.22em}\textcolor{green!50!black}{$\downarrow$}}
\newcommand{\upR}{\hspace{0.22em}\textcolor{red!80!black}{$\uparrow$}}
\newcommand{\dnR}{\hspace{0.22em}\textcolor{red!80!black}{$\downarrow$}}

\definecolor{primaryDark}{RGB}{30, 41, 59}      
\definecolor{accentColor}{RGB}{14, 116, 144}    
\definecolor{codeBg}{RGB}{248, 250, 252}        
\definecolor{codeFrame}{RGB}{226, 232, 240}     
\definecolor{codeComment}{RGB}{100, 116, 139}   
\definecolor{codeKeyword}{RGB}{194, 65, 12}     
\definecolor{codeString}{RGB}{15, 118, 110}     
\definecolor{codeInlineBg}{RGB}{241, 245, 249}  

\lstdefinestyle{bashStyle}{
    language=bash,
    backgroundcolor=\color{codeBg},
    basicstyle=\footnotesize\ttfamily\color{primaryDark},
    keywordstyle=\bfseries\color{codeKeyword},
    stringstyle=\color{codeString},
    commentstyle=\itshape\color{codeComment},
    numberstyle=\tiny\color{codeComment},
    showstringspaces=false,
    breaklines=true,
    breakatwhitespace=true,
    frame=single,
    framesep=5pt,
    xleftmargin=1pt,
    framexleftmargin=1pt,
    tabsize=2,
    captionpos=b
}

\begin{document}

\title{Do LLMs Understand Network Traces? A Comparative Study of Reasoning over Raw and Filtered Traffic Data\\

\thanks{Spanish Ministry of Science, Innovation and Universities (project VIGILANT, Ref. PID2024-161902OB-I00)}
}

\author{\IEEEauthorblockN{Irene Gosálvez-White, Roberto Magán-Carrión, Carlota de~la~Vega-Soriano and Pedro García-Teodoro}
\IEEEauthorblockA{\textit{Network Engineering \& Security Group - CITIC} \\
\textit{School of Computer Science and Telecommunication Engineering} \\
University of Granada\\
Granada, Spain \\
\{irenegwh,rmagan,carlotadelavega,pgteodor\}@ugr.es}
}

\maketitle

\begin{abstract}
The effective delivery and monitoring of ICT services rely heavily on network traffic analysis. However, interpreting complex packet captures remains a challenging task that requires substantial technical expertise. Large language models (LLMs) offer promising opportunities to automate network traffic investigation, yet their reliability in packet-level field extraction and flow-level sequence reconstruction remains an open research question. This study proposes a reproducible, question-answering-based evaluation methodology to assess the capabilities of LLMs in network trace analysis. Using real-world benign traffic from the CICIDS2017 dataset, we evaluate multiple models against a human-generated ground truth, comparing their performance on raw textual traffic dumps and protocol-filtered representations. Model responses are assessed through an automated LLM-as-a-judge pipeline, validated against human evaluations. The results demonstrate that protocol filtering significantly reduces prompt overhead while improving answer accuracy by reducing incorrect responses. These findings highlight the importance of protocol-aware traffic filtering as an effective optimization for improving the efficiency and reliability of LLM-assisted network management.
\end{abstract}


\section{Introduction}

Network traffic inspection plays a fundamental role in troubleshooting, performance diagnosis, and security incident investigation. These activities often require analysts to examine packet captures, interpret protocol fields, reconstruct communication sequences, and identify patterns that explain observed network behavior. Although established tools such as network analyzers and automated intrusion detection systems provide powerful capabilities for these tasks, interpreting complex traffic can still require substantial technical expertise, particularly due to the large volume of packets, the variety of protocols involved,  the relationships between packets across protocol layers, or when the analysis involves long packet sequences, unfamiliar protocols, or ambiguous network events.

Large language models (LLMs) have attracted considerable interest as potential assistants for network analysis \cite{Cui25}. Their ability to process textual information, follow instructions, summarize evidence, and generate explanations creates opportunities for more accessible and flexible traffic investigation. Rather than relying exclusively on predefined rules or specialized analysis pipelines, an LLM can be asked to explain a packet exchange, interpret protocol behavior, identify suspicious communication patterns, or answer questions about a packet capture.

Recent research has begun to explore the application of LLMs to networking and cybersecurity, including network operations, traffic representation, encrypted traffic analysis, and the interpretation or generation of network traffic \cite{Liu25,Fricke26,Nakamura26}. Although these developments suggest that LLMs may provide useful support for network analysts, their practical value depends on an important question: how effectively can these models understand network traffic, and how can that effectiveness be measured reliably?

Network traffic understanding comprises several complementary capabilities that should be evaluated separately. At the packet level, a model should correctly identify protocol fields, addresses, ports, flags, and other relevant attributes. At the flow level, it should relate individual packets to reconstruct communication sequences and interpret interactions between endpoints. At a higher level, it may need to explain connection failures, identify unusual behavior, or assess whether a trace supports a particular security hypothesis. This way, although existing research has investigated and demonstrated the potential of LLM-based for traffic classification and packet analysis, they also reveal performance differences across tasks and data formats \cite{Cui25,Paukkeri26,Ginige26}. An LLM may correctly identify a protocol but misinterpret the sequence of messages, overlook a retransmission, confuse a connection attempt with an established connection, or infer malicious intent from information that is insufficient to support such a conclusion. Similarly, a model may provide a technically convincing explanation that contains unsupported claims or contradicts details present in the packet capture.

As a consequence of the above, an effective evaluation should therefore cover multiple levels of analysis and use task-specific criteria to identify where each model performs well or poorly. For this reason, model outputs should be examined to determine whether the stated conclusions are supported by observable packets, fields, or communication patterns.
A comprehensive evaluation should consequently combine objective performance metrics with systematic assessment of generated explanations. 

In this work, we propose a methodology for evaluating LLMs through a question-answering (Q/A) task involving network traffic traces and protocols across the TCP/IP model. The main contributions of the work are:

\begin{itemize}
    \item \textbf{Reproducible evaluation of LLMs for network trace analysis:} A reproducible evaluation methodology is established to compare multiple LLMs on a common set of questions about network traffic traces, using reference answers generated through PyShark-based functions and consistent evaluation criteria.
    \item \textbf{Impact of protocol-aware preprocessing:} The effects of protocol-based filtering on answer accuracy are evaluated by comparing filtered traces with a general textual representation of the same trace, identifying the benefits and limitations of each approach.
    \item \textbf{Automated evaluation and error pattern analysis:} An evaluation pipeline is developed to automate response assessment and compare LLM-as-a-judge evaluations with human judgements. Recurring error patterns are also identified through manual analysis.
    \item \textbf{Effect of input size on real traffic:} The same models are evaluated on trace extracts of increasing size from a public traffic capture from the CICIDS2017 dataset~\cite{Sharafaldin18}. This analysis characterises how answer correctness changes as the amount of input data increases up to the model's context window.
\end{itemize}

The remainder of this paper is organised as follows. Section~\ref{sec:background} reviews related work in the field. Section~\ref{sec:methodology}  describes the introduced methodology, including the trace representations, question set, reference answers, and evaluation procedure. Section~\ref{sec:experimentalsetup} presents the experimental setup. Section~\ref{sec:results}  reports and discusses the results, and Section~\ref{sec:conclusions} concludes the paper and outlines directions for future work.

\section{Related Work}
\label{sec:background}

Network traffic analysis is an essential component of modern network management and cybersecurity. It supports the identification of applications and protocols, the diagnosis of communication failures, the detection of anomalous behavior, or the investigation of security incidents. Traditionally, these tasks have relied on protocol specifications, rule-based systems, statistical analysis, and the expertise of network administrators and security analysts. However, increasing network complexity, the diversity of connected devices, and the widespread adoption of encryption have made traffic analysis increasingly challenging. In response, artificial intelligence (AI) techniques have progressively complemented conventional approaches, enabling the identification of complex patterns in network traffic and supporting automated classification and threat detection. 

Early AI-based approaches to network traffic analysis primarily employed supervised and unsupervised ML algorithms to classify traffic, identify malicious activities, and detect deviations from normal network behavior. Traditional ML and DL models often require representative training datasets and task-specific optimization. Their performance may deteriorate when confronted with previously unseen traffic patterns, evolving attacks, or changes in the underlying network environment. These limitations have motivated research into more generalizable traffic representations and models capable of adapting to heterogeneous analysis tasks. 

Large language models (LLMs) has introduced new possibilities for AI-assisted network analysis. Unlike conventional classifiers designed primarily to map predefined features to labels, LLMs can interpret structured or textual descriptions, respond to natural-language questions, and generate explanations of their predictions. These capabilities may facilitate tasks such as protocol interpretation, packet inspection, traffic summarization, troubleshooting, and preliminary security investigation. Some works identify networking and security as relevant application areas for LLMs, while also discussing limitations related to scalability, explainability, privacy, and integration with existing network infrastructures \cite{Liu25,Cui25,Soto25,Enisoglu26}.

Similarly, although a number of technical proposals exist in the specialized literature regarding the use of LLMs for network-related tasks, like synthetic traffic generation, packet analysis, information extraction \cite{Kholgh23, Sharma23,Kan24,Wang24, Kwon25,Ginige26,Maatouk26}, the ability of general-purpose LLMs to interpret network traffic reliably remains an open question.

Three closely related studies explore the use of LLMs for answering questions about network traffic traces. \texttt{PCAP-LM} introduces a flow-centric textual representation that encodes packet-level behaviour and flow summaries into a compact format designed for LLM consumption~\cite{Marjou26}. Its evaluation on 5G/4G TLS 1.3 bulk-download traffic reports higher question-answering accuracy than a token-budget-matched prefix of \texttt{TShark} output, highlighting the importance of input representation when analysing network traces with LLMs. \texttt{TACO} solution combines an LLM-based agent, retrieval-augmented generation, and filtering tools to answer natural-language queries about real 5G core signalling traffic~\cite{Tothfalusi25}. 

However, a systematic and reproducible evaluation is still needed to determine how reliably LLMs can answer natural-language questions about multi-packet network traces, including tasks involving value extraction and counting, cross-packet aggregation, and traffic filtering. The effects of traffic representation, input size, and context window limitations also remain to be investigated. Dynamically preprocessing traffic traces according to the user's query could further support this intent-based approach.

Authors in \cite{Vico24} proposed a methodology for evaluating LLMs on text-based network traces using question-answering tasks, reference answers, and different prompting strategies. Building on this work, we extend the evaluation of LLMs for network trace analysis by developing a reproducible pipeline, presented below in Section~\ref{sec:methodology}, to systematically evaluate different models using a common set of questions and traces, examining the effects of traffic representation and input size, comparing automated response evaluations with human judgements, analysing recurring errors, and exploring an LLM-based agent capable of selecting and applying traffic filters according to the user's question. The complete project can be found at the public repository given in \cite{nesgRepo}.

\section{Methodology}
\label{sec:methodology}

In the previous context, we introduce in this section a reproducible methodology for evaluating LLMs on question-answering (Q/A) tasks for network traffic trace analysis. The methodology is designed to assess answer correctness across different network traffic traces, protocol-filtering criteria, among others conditions. The evaluation of how relevant is the effect on the system performance when only specific TCP/IP protocol stack layers are considered is also assessed. Answers are evaluated automatically using an LLM-as-a-judge approach, and its alignment with human evaluation 

\begin{figure}[t!]
    \centering
    \includegraphics[width=1\linewidth]{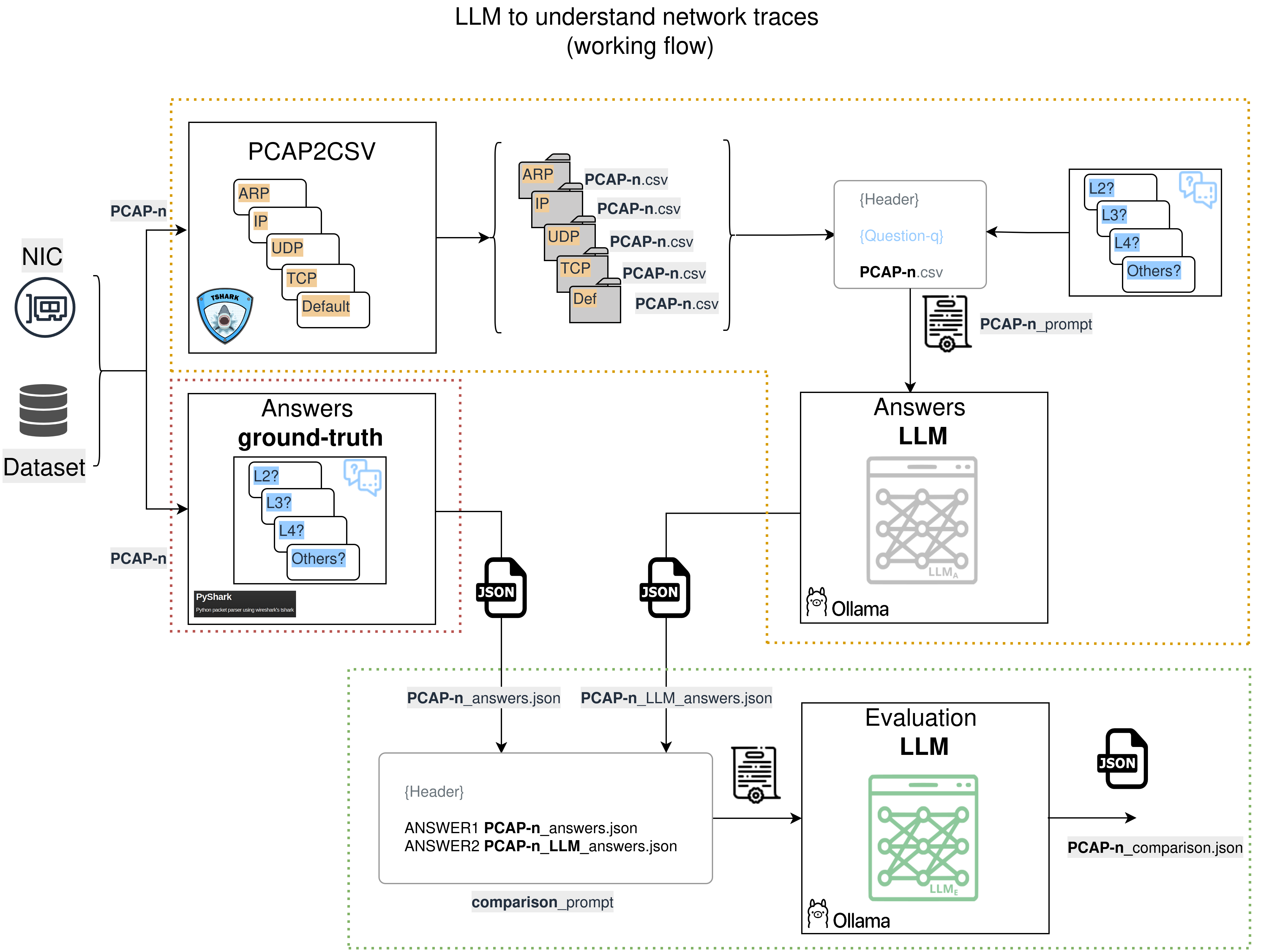}
    \caption{Overview of the proposed workflow for evaluating LLMs in network trace analysis.}
    \label{fig:workflow}
\end{figure}

As illustrated in Figure~\ref{fig:workflow}, the methodology's workflow consists of the following stages:

\begin{enumerate}
     \item \textbf{Network data:} Selection of network traces in \texttt{PCAPNG} format as input of the system. They may either originate from benchmark network datasets or be captured directly via the network interface card (NIC) for offline analysis. 
     \item \textbf{Question-answer dataset generation:} Definition of a question-answer dataset (ground truth) to assess the ability of LLMs to extract, count, aggregate, and interpret information from the input network traffic data.  
     \item \textbf{Data preprocessing:} Converting \texttt{PCAPNG} network traces into \texttt{CSV} files using \texttt{TShark}\footnote{\url{https://tshark.dev/}} tool in two ways: a) the default conversion, and b) the protocol-filter approach conversion. The resulting \texttt{CSV} contents are part of the generated prompts for feeding the LLMs.
     \item \textbf{Prompt and answer generation:} Prompt generation to feed the under-evaluation LLM whose answers will be compared against the questions-answer dataset.
    \item \textbf{Performance evaluation:} Comparison and classification of LLM-generated answers against the ground-truth both human-driven and following a LLM-as-a-judge approach. For that, performance metrics will be defined for answer correctness evaluation with or without protocol filtering.
    
\end{enumerate}

\subsection{Network data}

In order to assess how LLMs understand network traffic, the CICIDS2017~\cite{Sharafaldin18} dataset is used, built up by the Canadian Institute of Cyber Security. The dataset comprises both normal traffic and well known attacks such as Brute Force Attack, DoS/DDoS, etc. 

\subsection{Question-answer dataset generation}

Assessing systems in general and AI-based systems in particular involves the generation of datasets to validate their outcomes and to be confident about the performance they are offering. In this context, these datasets are widely-known as the ground truth which will be compared with the output provided by the AI-based system. In this paper, a set of questions and their corresponding answers are developed by the authors as experts in the field. Therefore, each question-answer pairs are computed automatically from the original \texttt{PCAPNG} traces using question-specific \texttt{PyShark} functions. Each function extracts the required information from the network traffic sample and formats the answer according to a predefined template; the question-answer pairs are stored in \texttt{JSON} files and are available at the project repository. 

Due to it relevance in networking ARP, IPv4, TCP and UDP protocols from the link, network and transport layers of the TCP/IP stack have been chosen. Nineteen (19) are the questions designed, where 3 refer to ARP, 4 to IPv4, 9 to TCP and 3 to UDP. They range from direct field extraction (e.g., unique MAC or IP addresses, or well-known destination ports) to counting, aggregation and interpretation across packets (e.g., frequent ports, private and public addresses with their packet counts, and ports associated with specific connection states or patterns, such as connections ending with FIN or RST, completed or failed 
-way handshakes, or repeated SYN packets without a response). The complete question-answer dataset is available at the project repository.

\subsection{Data preprocessing}

LLMs interpret natural language. In fact, they were built to be fed with text. Because of that, the original \texttt{PCAPNG} traces needs to be converted. To this end, \texttt{TShark} was used to convert \texttt{PCAPNG} files into \texttt{CSV} format, which is more readily interpretable by LLMs than the original format. 

Two ways of preprocessing are carried out, hereafter called views:

\begin{itemize}
    \item \textbf{Default view:} In this view, each line represents a packet having the frame number, timestamp, IP addresses, protocol, length and a brief description of it as contents. It covers all the traffic and protocols of the trace, but does not expose all packet fields.
    \item \textbf{Protocol-filter views:} For each protocol involved in the question set, predefined protocol packet fields are chosen. These fields have been carefully selected to maximize the information we can extract from that protocol. 
\end{itemize}

Considering these two ways of preprocessing we will able to evaluate the impact on the system performance first, considering all the captured traffic trace, difficult to manage due to the inherent nature of the network traffic, and second, how having concise traffic information could reduce the complexity of the information to be managed keeping the system performance or even increasing it.

More details about the considered protocols and per view fields are described in Section~\ref{sec:experimentalsetup}.

\subsection{Prompt and answer generation}

After the preprocessing stage,  each converted network traffic trace is provided to the LLM together with the questions defined in the question-answer dataset. For that, a prompt is generated following a zero-shot approach with the structure shown in Figure~\ref{box:prompt_structure}. For each question we record the answer together with its input and output token counts.

\begin{figure}[!b]
    \centering
    \begin{footnotesize}
    \begin{tcolorbox}[title=Prompt structure, 
    colback=gray!5!white, 
    colframe=gray!75!black]
        \textbf{1. Role and task:}
        Defines the role (network traffic analyst) and the task of answering a question about the trace. \\
        \textbf{2. Data:}
        The network trace: default view (one-line column description) or protocol-filtered view (with its header row). \\
        \textbf{3. Question:}
        One question of the dataset. \\
        \textbf{4. Guardrails and output requirements:}
        Constraints on the response (use only trace information --networking knowledge solely to interpret it
        --; \texttt{None} if nothing matches; concise answer in the requested format, with no preamble.).
    \end{tcolorbox}
    \end{footnotesize}
\caption{Detailed prompt structure and main sections.}
    \label{box:prompt_structure}
\end{figure}

\subsection{Performance evaluation}
\label{sec:performance}

Human-driven answer evaluation is a highly time-consuming and exhaustive process that heavily depends on the evaluator's expertise, while also remaining susceptible to human error. Moreover, simple automatic evaluation methods have limitations when assessing answers that combine natural language explanations with exact technical values: string-based methods struggle with linguistic variability, whereas semantic similarity metrics may overlook discrepancies in technical values (e.g., IP addresses or port numbers). Therefore, an LLM-as-a-judge approach has been used, which compares each answer with the ground truth and assigns one of following three labels: \textit{Correct} \textbf{(C)}, if the answer matches the reference, with no missing, incorrect or additional elements; \textit{Partial}\textbf{ (P)}, if it identifies part of the reference but omits elements or adds incorrect ones; and \textit{Incorrect} \textbf{(I)}, if it identifies none of it. Differences in order, formatting and duplicates are ignored.

For the LLM-as-a-judge approach, the judge is instructed with an explicit rubric that defines the classification criteria and the equivalent answer formats (rule augmentation), worked examples of each label (demonstration) and a brief justification before the label~\cite{b14}. To reduce sensitivity to the answer order, each answer is evaluated twice with the positions of the reference and the candidate swapped (swapping operation). If the two evaluations disagree, the stricter label is kept and the case is recorded as an inconsistency.

\vspace*{0.1cm}
The results are reported and grouped by model, data preprocessing technique used and, where applicable, input size, using the following metrics:

\begin{itemize}
    \item \textbf{Answer correctness:} Percentages of answers labelled C, P and I.
    \item \textbf{Effect of filtering:} Differences in these percentages between the default and protocol-filtered views (in percentage points) and, per question, whether the label improves, remains unchanged or worsens. 
    \item \textbf{Resource consumption:} Input tokens per prompt and the token reduction achieved by filtering.
    \item \textbf{Judge reliability:} Agreement with the manual assessments, measured by percentage agreement, Cohen's kappa and a weighted kappa~\cite{cohen1960, cohen1968} that treats the labels as ordinal ($\mathrm{C}>\mathrm{P}>\mathrm{I}$).
\end{itemize}

\section{Experimental Setup}
\label{sec:experimentalsetup}

The specific experimental setup to evaluate the capabilities of generic LLMs to understand network traffic is presented here. As previouly indicated, the CICIDS2017 dataset is used for that. Two kinds of network traffic, in \texttt{PCAP} format, are involved in this dataset: benign and attack network traffic. In the first one, there are no malicious actions at all and will be used in this work, specifically Monday 3rd of July 2017 which is the only one without any kind of attacks. More specifically, the first 50,000 packets of benign traffic captured on 3rd July are extracted for the experiments. 

It is also worth mentioning the execution of five runs of the complete methodology over the network dataset, in order to obtain statistical result related figures.

\subsection{Data preprocessing}

As introduced in Section~\ref{sec:methodology}, network traffic data is converted into text (\texttt{CSV}) views: default and protocol-filter approach. For that, \texttt{TShark} command line tool is used. In the following, both data preprocessing approaches  illustrative examples of each conversion are provided.

\subsubsection{Default view}
This is the default conversion as shown in the following listing: 
\begin{lstlisting}[
%caption={Default view \texttt{TShark} command.},
label={lst:default_view}]
$ tshark -n -r capture.pcapng -T tabs
\end{lstlisting}
    
\subsubsection{Protocol-filter views}
One view is generated for each protocol addressed by the questions using a protocol-specific filter. ARP, IPv4, TCP and UDP protocols are used. Each view retains the frame number, the relative timestamp and the header fields of its protocol (e.g., ports, sequence and acknowledgement numbers, flags and window size for TCP). The IPv4 protocol name is added alongside its numeric value, and TCP flags are represented by their names. The TCP and UDP views exclude ICMP and ICMPv6 packets, whose error messages quote the headers of the original packet, to avoid mistaking them for fields of the outer packet.

An example of the IPv4-based filter conversion is as follows:
\begin{lstlisting}[
%caption={Protocol-filtered (IPv4) view \texttt{TShark} command.},
label={lst:protocol_view}]
$ tshark  -n -r capture.pcapng -Y "ip" \
    -T fields -E header=y -E separator=, \
    -E occurrence=f \
    -e frame.number -e frame.time_relative \
    -e ip.version -e ip.hdr_len -e ip.dsfield \
    -e ip.len -e ip.id -e ip.flags \
    -e ip.frag_offset -e ip.ttl -e ip.proto \
    -e ip.checksum -e ip.src -e ip.dst
\end{lstlisting}
 
\subsection{LLM setup}

A total of four different LLMs belonging to two well-established model families (Llama and Qwen) were evaluated, with Llama 3.3 70B specifically serving as the judge model. Experiments are conducted locally using Ollama server v0.33.3 on a system equipped with an NVIDIA GB10 GPU, 121 GiB of RAM, and a 20-core ARM CPU. All generation models were configured with a temperature of 0 and a maximum output length of 1,024 tokens. Unless explicitly overridden, the context window is set to the maximum length reported by Ollama for each model. Table~\ref{tab:models} summarizes the evaluated models, their context windows, and their quantization.

\begin{table}[!b]
\caption{Evaluated models, context windows and quantisation.}
\label{tab:models}
\centering
\small
\setlength{\tabcolsep}{3pt}
\begin{tabularx}{\columnwidth}{@{}l r r l >{\raggedright\arraybackslash}X@{}}
\hline
\textbf{Model} & \textbf{Parameters} & \textbf{Context} & \textbf{Quant.} & \textbf{Role} \\
\hline
Llama 3.1& 8B  & 131,072 & Q4\_K\_M & Answer generation \\
Llama 3.1& 70.6B & 131,072 & Q4\_K\_M & Answer generation \\
Qwen 2.5& 7.6B  & 32,768  & Q4\_K\_M & Answer generation \\
Qwen 2.5& 72.7B & 32,768  & Q4\_K\_M & Answer generation \\
Llama 3.3& 70.6B & 131,072 & Q4\_K\_M & LLM-as-a-judge \\
\hline
\end{tabularx}
\end{table}

\subsection{Data sampling and experimental evaluation}

For each of the five runs executed, a contiguous chunk of packets, large enough to contain the largest level of every model, is selected starting at a random packet (the seed is the run number, so the starting packet differs between runs, while all models share the same chunk in each run). The chunk is converted into the default view and its tokens are counted with the tokenizer of each model.

To examine how answer correctness changes as more packet data is provided and less of the context window remains available to the model, the first $N$ packets of each chunk are selected so that the trace occupies an increasing proportion of the available input budget (50, 75, and 100\,\%). This budget is calculated by substracting from the model's context window the tokens required for the prompt excluding the trace data (see Fig.~\ref{box:prompt_structure}), the longest question, and the maximum output length. The latter is limited to prevent excessively long generations (e.g., looping responses) from causing inference timeouts.

At a input budget of 100\,\%, the complete prompt approaches the context window limit while still fitting within it. A 125\,\% budget is also generated but excluded from the analysis because the inference server truncates prompts exceeding the context window to half of the configured size, preventing the models from receiving the complete input. As the available input budget depends on the context window and tokenizer, $N$ differs between model families. In the first run, the three levels contain 1,087, 1,549, and 2,046 packets for Llama~3.1, and 183, 269, and 363 packets for Qwen~2.5. The protocol-filtered views are generated from the same packet subsets. For each run, the four models are evaluated on the 19 questions using the three input sizes and both data representations. This results in $19\text{ questions}\times4\text{ models}\times3\text{ input sizes}\times5\text{ runs}=1{,}140$ generated answers per representation (2,280 in total).

To assess the agreement between the LLM-as-a-judge and human evaluation, a subset of answers generated from captured network traces in a basic and controlled network environment (see Table~\ref{tab:basic_traces}) is manually assessed. This subset comprises 608 answers from the first run $1\text{ run}\times4\text{ models}\times4\text{ traces}\times2\text{ representations}\times19\text{ questions}$, evaluated by a single annotator against the corresponding ground-truth answers, without access to the LLM-as-a-judge labels.

\begin{table}[!b]
\caption{Basic environment: type of traffic and number of packets for the default view and per protocol-filter view.}
\label{tab:basic_traces}
\centering
\footnotesize
\setlength{\tabcolsep}{3pt}
\renewcommand{\arraystretch}{1.2}

\begin{tabularx}{\columnwidth}{@{} m{0.8cm} >{\centering\arraybackslash}X r r r r r @{}}
\hline
 & & \multicolumn{5}{c}{\textbf{Packets}} \\
\cline{3-7}
\textbf{Trace} & \textbf{Traffic} & Default & ARP & IP & TCP & UDP \\
\hline
T1 & Host initialisation (neighbour discovery, address resolution, connectivity checks) & 23 & 7 & 6 & 0 & 0 \\
T2 & Connectivity tests, DNS, IP fragmentation, failed TCP connections & 29 & 0 & 29 & 7 & 6 \\
T3 & Nmap-based TCP SYN and UDP scan, ICMP from closed ports & 32 & 4 & 27 & 13 & 8 \\
T4 & Host discovery and port scanning using ARP, UDP, TCP, and ICMP, including TCP three-way handshake completion & 34 & 4 & 30 & 16 & 8 \\
\hline
\end{tabularx}
\end{table}

\begin{table*}
\centering
\caption{Evaluation results for the CICIDS2017 benign scenario across portions of input tokens (50\%, 75\%, 100\%), protocol layers (L2: ARP, L3: IPv4, L4: TCP/UDP) and views (Default vs.\ Protocol-filtered, the last one shaded). \underline{Quantitative overview}: percentage of Correct (C), Partial (P) and Incorrect (I) answers aggregated over 5 evaluation runs; for the filtered view, $\uparrow$ denote an increase, $\downarrow$ a decrease and  $\equiv$ equivalence with respect to the default view. For C and I, \textcolor{green}{green} indicates an improvement and \textcolor{red}{red} a deterioration (arrows for P are not colored, as a change in P cannot be unambiguously classified). \underline{Qualitative overview}: transition score from the default to the filtered view (+2 for I$\rightarrow$C; +1 for I$\rightarrow$P and P$\rightarrow$C; opposite sign if the label worsens), scaled to $\pm$100\,\%; \textcolor{green}{\checkmark{}} indicates improvement, and \textcolor{red}{$\times$} deterioration.}
\label{tab:evaluacion_desglosada}
\footnotesize
\setlength{\tabcolsep}{2.2pt} 
\renewcommand{\arraystretch}{1.12}
\begin{tabular}{|l|c|l||c|c|c||c|c|c||c|c|c||c|c|c||c|}
\hline
\multirow{2}{*}{\textbf{Model}} & \multirow{2}{*}{\textbf{Portion (\%)}} & \multirow{2}{*}{\textbf{View}} & \multicolumn{3}{c||}{\textbf{L2 (ARP)}} & \multicolumn{3}{c||}{\textbf{L3 (IP)}} & \multicolumn{3}{c||}{\textbf{L4 (TCP/UDP)}} & \multicolumn{3}{c||}{\textbf{\shortstack{\rule{0pt}{2.4ex}Quantitative\\overview}}} & \multirow{2}{*}{\textbf{\shortstack{\rule{0pt}{2.4ex}Qualitative\\overview}}} \\ \cline{4-15} 
 & & & \textbf{C} & \textbf{P} & \textbf{I} & \textbf{C} & \textbf{P} & \textbf{I} & \textbf{C} & \textbf{P} & \textbf{I} & \textbf{C} & \textbf{P} & \textbf{I} & \\ \hline

\multirow{6}{*}{Qwen 2.5 7B} 
 & \multirow{2}{*}{50\%} & Def. & 60 & 13.3 & 26.7 & 5 & 45 & 50 & 51.7 & 6.7 & 41.7 & 43.2 & 15.8 & 41.1 & \multirow{2}{*}{\imp{4.2}} \\ \cline{3-15}
 &                      & \cellcolor{gray!15}Fil. & \cellcolor{gray!15}53.3\dnR & \cellcolor{gray!15}$20\uparrow$ & \cellcolor{gray!15}$26.6\equiv$ & \cellcolor{gray!15}15\upG & \cellcolor{gray!15}$60\uparrow$ & \cellcolor{gray!15}25\dnG & \cellcolor{gray!15}50\dnR & \cellcolor{gray!15}$13.3\uparrow$ & \cellcolor{gray!15}36.7\dnG & \cellcolor{gray!15}$43.1\equiv$ & \cellcolor{gray!15}$24.2\uparrow$ & \cellcolor{gray!15}32.6\dnG & \\ \cline{2-16}
 & \multirow{2}{*}{75\%} & Def. & 46.7 & 13.3 & 40 & 5 & 45 & 50 & 21.6 & 21.7 & 56.7 & 22.1 & 25.3 & 52.6 & \multirow{2}{*}{\imp{11.6}} \\ \cline{3-15}
 &                      & \cellcolor{gray!15}Fil. & \cellcolor{gray!15}$46.7\equiv$ & \cellcolor{gray!15}$26.7\uparrow$ & \cellcolor{gray!15}26.6\dnG & \cellcolor{gray!15}15\upG & \cellcolor{gray!15}$75\uparrow$ & \cellcolor{gray!15}10\dnG & \cellcolor{gray!15}23.3\upG & \cellcolor{gray!15}$35\uparrow$ & \cellcolor{gray!15}41.7\dnG & \cellcolor{gray!15}25.3\upG & \cellcolor{gray!15}$42.1\uparrow$ & \cellcolor{gray!15}32.6\dnG & \\ \cline{2-16}
 & \multirow{2}{*}{100\%} & Def. & 60 & 6.7 & 33.3 & 5 & 40 & 55 & 20 & 18.3 & 61.7 & 23.2 & 21.1 & 55.8 & \multirow{2}{*}{\imp{6.3}} \\ \cline{3-15}
 &                      & \cellcolor{gray!15}Fil. & \cellcolor{gray!15}46.7\dnR & \cellcolor{gray!15}$26.7\uparrow$ & \cellcolor{gray!15}26.6\dnG & \cellcolor{gray!15}20\upG & \cellcolor{gray!15}$50\uparrow$ & \cellcolor{gray!15}30\dnG & \cellcolor{gray!15}16.7\dnR & \cellcolor{gray!15}$33.3\uparrow$ & \cellcolor{gray!15}50\dnG & \cellcolor{gray!15}22.1\dnR & \cellcolor{gray!15}$35.8\uparrow$ & \cellcolor{gray!15}42.1\dnG & \\ \hline\hline

\multirow{6}{*}{Qwen 2.5 72B} 
 & \multirow{2}{*}{50\%} & Def. & 86.7 & 13.3 & 0.0 & 25 & 70 & 5 & 65 & 8.3 & 26.7 & \textbf{\textcolor{green}{60}} & 22.1 & \textbf{\textcolor{green}{17.9}} & \multirow{2}{*}{\imp{2.1}} \\ \cline{3-15}
 &                      & \cellcolor{gray!15}Fil. & \cellcolor{gray!15}73.3\dnR & \cellcolor{gray!15}$13.3\equiv$ & \cellcolor{gray!15}13.3\upR & \cellcolor{gray!15}45\upG & \cellcolor{gray!15}$55\downarrow$ & \cellcolor{gray!15}0\dnG & \cellcolor{gray!15}60\dnR & \cellcolor{gray!15}$23.3\uparrow$ & \cellcolor{gray!15}16.7\dnG & \cellcolor{gray!15}58.9\dnR & \cellcolor{gray!15}$28.4\uparrow$ & \cellcolor{gray!15}12.6\dnG & \\ \cline{2-16}
 & \multirow{2}{*}{75\%} & Def. & 80 & 13.3 & 6.7 & 30 & 60 & 10 & 38.3 & 18.4 & 43.3 & 43.1 & 26.3 & 30.5 & \multirow{2}{*}{\imp{10.5}} \\ \cline{3-15}
 &                      & \cellcolor{gray!15}Fil. & \cellcolor{gray!15}66.7\dnR & \cellcolor{gray!15}$20\uparrow$ & \cellcolor{gray!15}13.3\upR & \cellcolor{gray!15}45\upG & \cellcolor{gray!15}$55\downarrow$ & \cellcolor{gray!15}0\dnG & \cellcolor{gray!15}51.7\upG & \cellcolor{gray!15}$21.7\uparrow$ & \cellcolor{gray!15}26.7\dnG & \cellcolor{gray!15}52.6\upG & \cellcolor{gray!15}$28.4\uparrow$ & \cellcolor{gray!15}18.9\dnG & \\ \cline{2-16}
 & \multirow{2}{*}{100\%} & Def. & 86.7 & 13.3 & 0 & 30 & 55 & 15 & 30 & 28.3 & 41.7 & 39 & 31.6 & 29.5 & \multirow{2}{*}{\imp{5.3}} \\ \cline{3-15}
 &                      & \cellcolor{gray!15}Fil. & \cellcolor{gray!15}66.7\dnR & \cellcolor{gray!15}$20\uparrow$ & \cellcolor{gray!15}13.3\upR & \cellcolor{gray!15}40\upG & \cellcolor{gray!15}$60\uparrow$ & \cellcolor{gray!15}0\dnG & \cellcolor{gray!15}43.3\upG & \cellcolor{gray!15}$18.3\downarrow$ & \cellcolor{gray!15}38.4\dnG & \cellcolor{gray!15}46.3\upG & \cellcolor{gray!15}$27.4\downarrow$ & \cellcolor{gray!15}26.3\dnG & \\ \hline\hline

\multirow{6}{*}{Llama 3.1 8B} 
 & \multirow{2}{*}{50\%} & Def. & 20 & 13.3 & 66.7 & 5 & 70 & 25 & 3.3 & 36.7 & 60 & 6.3 & 40 & 53.7 & \multirow{2}{*}{\boldmath\bfseries\textcolor{red}{\wor{1.1}}} \\ \cline{3-15}
 &                      & \cellcolor{gray!15}Fil. & \cellcolor{gray!15}6.7\dnR & \cellcolor{gray!15}$26.7\uparrow$ & \cellcolor{gray!15}$66.7\equiv$ & \cellcolor{gray!15}$5\equiv$ & \cellcolor{gray!15}$65\downarrow$ & \cellcolor{gray!15}30\upR & \cellcolor{gray!15}$3.3\equiv$ & \cellcolor{gray!15}$38.3\uparrow$ & \cellcolor{gray!15}58.3\dnG & \cellcolor{gray!15}4.2\dnR & \cellcolor{gray!15}$42.1\uparrow$ & \cellcolor{gray!15}$53.7\equiv$ & \\ \cline{2-16}
 & \multirow{2}{*}{75\%} & Def. & 0 & 20 & 80 & 5 & 65 & 30 & 3.3 & 28.3 & 68.3 & \textbf{\textcolor{red}{3.2}} & 34.7 & 62.1 & \multirow{2}{*}{\imp{6.3}} \\ \cline{3-15}
 &                      & \cellcolor{gray!15}Fil. & \cellcolor{gray!15}$0\equiv$ & \cellcolor{gray!15}$40\uparrow$ & \cellcolor{gray!15}60\dnG & \cellcolor{gray!15}10\upG & \cellcolor{gray!15}$50\downarrow$ & \cellcolor{gray!15}40\upR & \cellcolor{gray!15}8.3\upG & \cellcolor{gray!15}$35\uparrow$ & \cellcolor{gray!15}56.7\dnG & \cellcolor{gray!15}7.4\upG & \cellcolor{gray!15}$38.9\uparrow$ & \cellcolor{gray!15}53.7\dnG & \\ \cline{2-16}
 & \multirow{2}{*}{100\%} & Def. & 13.3 & 20 & 66.7 & 0 & 40 & 60 & 3.3 & 23.3 & 73.3 & 4.2 & 26.3 & \textbf{\textcolor{red}{69.5}} & \multirow{2}{*}{\imp{5.8}} \\ \cline{3-15}
 &                      & \cellcolor{gray!15}Fil. & \cellcolor{gray!15}$13.3\equiv$ & \cellcolor{gray!15}$53.3\uparrow$ & \cellcolor{gray!15}33.3\dnG & \cellcolor{gray!15}$0\equiv$ & \cellcolor{gray!15}$45\uparrow$ & \cellcolor{gray!15}55\dnG & \cellcolor{gray!15}6.7\upG & \cellcolor{gray!15}$25\uparrow$ & \cellcolor{gray!15}68.3\dnG & \cellcolor{gray!15}6.3\upG & \cellcolor{gray!15}$33.7\uparrow$ & \cellcolor{gray!15}60\dnG & \\ \hline\hline

\multirow{6}{*}{Llama 3.1 70B} 
 & \multirow{2}{*}{50\%} & Def. & 46.7 & 26.7 & 26.7 & 10 & 85 & 5 & 16.7 & 35 & 48.3 & 20 & 44.2 & 35.8 & \multirow{2}{*}{\imp{6.8}} \\ \cline{3-15}
 &                      & \cellcolor{gray!15}Fil. & \cellcolor{gray!15}40\dnR & \cellcolor{gray!15}$13.3\downarrow$ & \cellcolor{gray!15}46.7\upR & \cellcolor{gray!15}35\upG & \cellcolor{gray!15}$60\downarrow$ & \cellcolor{gray!15}$5\equiv$ & \cellcolor{gray!15}26.7\upG & \cellcolor{gray!15}$35\equiv$ & \cellcolor{gray!15}38.3\dnG & \cellcolor{gray!15}30.5\upG & \cellcolor{gray!15}$36.8\downarrow$ & \cellcolor{gray!15}32.6\dnG & \\ \cline{2-16}
 & \multirow{2}{*}{75\%} & Def. & 13.3 & 33.3 & 53.3 & 0 & 80 & 20 & 15 & 35 & 50 & 11.6 & 44.2 & 44.2 & \multirow{2}{*}{\imp{9.5}} \\ \cline{3-15}
 &                      & \cellcolor{gray!15}Fil. & \cellcolor{gray!15}33.3\upG & \cellcolor{gray!15}$33.3\equiv$ & \cellcolor{gray!15}33.3\dnG & \cellcolor{gray!15}25\upG & \cellcolor{gray!15}$65\downarrow$ & \cellcolor{gray!15}10\dnG & \cellcolor{gray!15}$15\equiv$ & \cellcolor{gray!15}$43.3\uparrow$ & \cellcolor{gray!15}41.7\dnG & \cellcolor{gray!15}20\upG & \cellcolor{gray!15}$46.3\uparrow$ & \cellcolor{gray!15}33.7\dnG & \\ \cline{2-16}
 & \multirow{2}{*}{100\%} & Def. & 13.3 & 26.6 & 60 & 0 & 25 & 75 & 11.7 & 25 & 63.3 & 9.5 & 25.2 & 65.2 & \multirow{2}{*}{{\boldmath\bfseries\textcolor{green}{\imp{15.3}}}} \\ \cline{3-15}
 &                      & \cellcolor{gray!15}Fil. & \cellcolor{gray!15}26.7\upG & \cellcolor{gray!15}$53.3\uparrow$ & \cellcolor{gray!15}20\dnG & \cellcolor{gray!15}5\upG & \cellcolor{gray!15}$65\uparrow$ & \cellcolor{gray!15}30\dnG & \cellcolor{gray!15}$11.7\equiv$ & \cellcolor{gray!15}$43.3\uparrow$ & \cellcolor{gray!15}45\dnG & \cellcolor{gray!15}12.6\upG & \cellcolor{gray!15}$49.5\uparrow$ & \cellcolor{gray!15}37.9\dnG & \\ \hline
\end{tabular}
\end{table*}

\begin{figure*}[t!]
    \centering
    \includegraphics[width=0.46\textwidth]{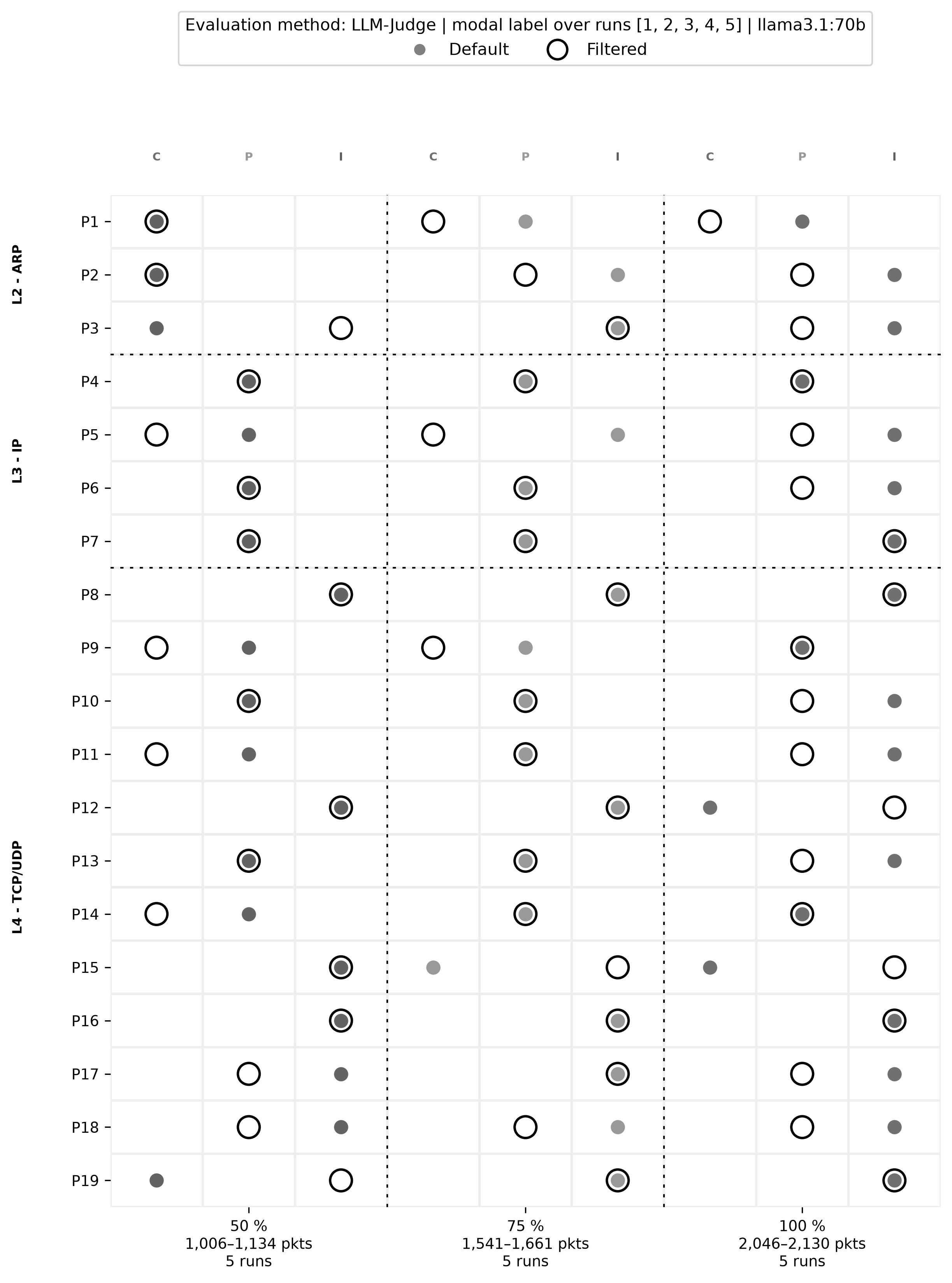}\hfill
    \includegraphics[width=0.46\textwidth]{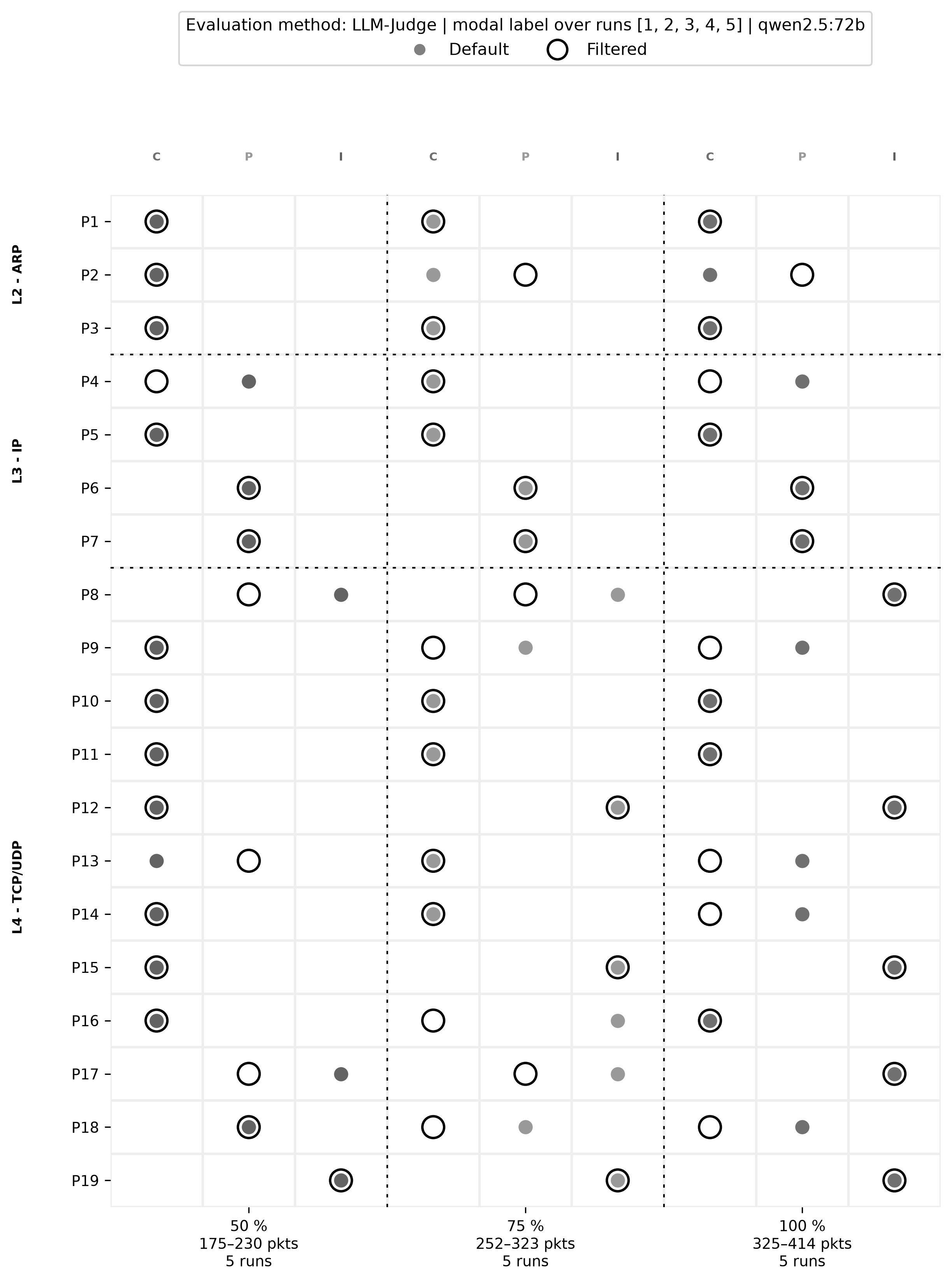}
    \caption{Modal label (C, P or I) of each question across the five runs of CICIDS2017, for Llama~3.1~70B (left) and Qwen~2.5~72B (right) models, at three portions of inputs packets (chunks). Ties between equally frequent labels are resolved in favor of the better label (C $>$ P $>$ I). P1--P3 address ARP protocol while P4--P7 IPv4 and P8--P19 TCP/UDP, respectively.}
    \label{fig:dotplot-cicids}
\end{figure*}

\section{Results and discussion}
\label{sec:results}

The main findings and conclusions drawn from the experimental results are summarised below, based on the performance metrics presented in Section \ref{sec:performance}.

\vspace*{0.1cm}
$\bullet$ {\bf Judge reliability}.  The judge agrees on 95.1\,\% of the 608 answers evaluated by the human annotator (unweighted Cohen's $\kappa=0.92$; weighted Cohen's $\kappa=0.94$). The weighted coefficient accounts for the ordinal nature of the labels, assigning lower penalties to disagreements between adjacent categories (e.g., Correct and Partial) than between more distant ones (e.g., Correct and Incorrect). In 29 of the 30 disagreements, the LLM was stricter than the human annotator. These results support the use of the same judge to evaluate the CICIDS2017 dataset samples.

\vspace*{0.1cm}
$\bullet$ {\bf Answer correctness}. On the CICIDS2017 dataset (see Table~\ref{tab:evaluacion_desglosada}), the comparison between families is not symmetrical: as the budget scales with the context window, Llama~3.1 receives about five times more packets than Qwen~2.5 (1,006--2,130 vs. 175--414 packets across runs and levels), so the differences reflect both the model and the amount of data.

Considering the quantitative overview in the table, the Qwen~2.5 model family obtains a higher answer correctness than Llama~3.1 model families, no matter the considered portion or view. Specifically, and regarding the default view, Qwen~2.5~72B results range from $C=39\,\%$ to $C=60\,\%$ while Qwen~2.5~7B results range from $C=22.1\,\%$ to $C=43.2\,\%$. In case of Llama~3.1~70B it ranges from $C=20\,\%$ to $C=9.5\,\%$ while Llama~3.1~8B ranges from $C=3.2\,\%$ to $C=6.3\,\%$. It can be observed that the higher the portion of input tokens is, the lower the answer correctness. Conversely, the incorrectness increases with the portion of input tokens, being the Llama~3.1~8B model the worse with $I=69.5\,\%$ while the Qwen~2.5~72B is the better with $I=17.9\,\%$.


{\vspace*{0.1cm}
$\bullet$ {\bf Effect of filtering}. For the easy understanding of this effect, the transition score was devised and represented in the  qualitative overview column of the table. 11 out of 12 measurements are positive ranging from $+2.1$ to $+15.3$, which means the protocol-filter view improves the LLM performance in understanding network traces as expected. Llama~3.1~8B model was the only one with a negative score of $-1.1$, of the 95 paired answers for this portion, 11 improved and 10 worsened, while 74 remained unchanged.

Specifically, and in regards to the quantitative overview column of the table, we can conclude that the protocol filtering approach behaviors as expected: it improves both the correctness and incorrectness in almost all scenarios. It can be there, in the Table \ref{tab:evaluacion_desglosada}, interpreted as green upward and downward arrows for correctness and incorrectness respectively. Few are the scenarios where the filtering approach does not significantly improve the correctness. They are for Qwen~2.5~72B and Llama~3.1~8B modes with a decrement of a $+0.7$ and $+2.1$ respectively. However, the partially correct answers increase while the incorrect answers decrease.

\vspace*{0.1cm}
$\bullet$ {\bf Resource consumption}. Complementing its positive impact on correctness, protocol-level filtering yields a pronounced effect on resource consumption and layer-, protocol-specific compaction, drastically compressing prompt length across all evaluated scenarios. The filtered representation achieves an average token reduction between $57.7\,\%$ and $64.4\,\%$, with median reductions reaching up to $79.8\,\%$ in Llama~3.1 and $68.0\,\%$ in Qwen~2.5 models families respectively. In absolute terms, the mean prompt length for Qwen~2.5~72B decreases from 15,792--31,397 input tokens in the default view down to 5,251--10,449 tokens under filtering (and from 64,447--128,778 down to 22,027--45,229 tokens in Llama~3.1~70B).

\subsection{Failure patterns}
Figure~\ref{fig:dotplot-cicids} shows a question-by-question breakdown of the results for \mbox{Llama~3.1~70B} and \mbox{Qwen~2.5~72B} models since they were the best as mentioned before. The layer-, protocol-specific questions are represented in the Y-axis in the figure while the used portions, measured in network packets, in the X-axis. Filled dots represent the modal computation across 5 runs of the performance in terms of C, P and I default view results while empty circles represent the same but in the case of protocol-filter view. The results reveal clear differences between question types. Simple lookups, such as identifying IP or MAC addresses (questions P1 to P5), remain relatively stable as the portions of network packets increases. In contrast, questions involving packet counting or state tracking (question P8 to P19 in L4-TCP/UDP) become much harder to answer in the default view. Interested readers are kindly invited to see the complete question-answer dataset in~\cite{nesgRepo} 

Across both model families, four recurring failure patterns are observed. First, models have difficulty handling missing data and empty results. When the asked protocol is absent from the chunk, the filtered view is empty and the expected answer is \texttt{None}. Three of the four models often ignore this instruction and answer with the column names of the empty table or with invented values, so the score worsens, whereas for the questions whose reference answer has content filtering improves the four models. This is particularly noticeable in \mbox{Llama~3.1~8B}, where empty filtered tables or queries about absent public IPs often lead to hallucinated service ports and network flows instead of the expected \texttt{None} response. 

Second, packet counting remains a major challenge regardless of model size. Frequency-based queries, such as counting packets in each direction for an IP or ranking the most common ports, often produce results that differ from the ground truth.

Third, models struggle with deduplication and aggregation, sometimes repeating the same entries or mixing source and destination counts.

Finally, tracking events across multiple packets is still unreliable. For example, models sometimes reverse client--server roles during TCP RST teardowns or report completed three-way handshakes in scan traffic, even when connections were refused or ignored. Some of these difficulties in tracking TCP sessions are also related to a limitation in our filtering design. The current transport-layer view only includes TCP fields and excludes IPv4 addresses, which are needed to distinguish concurrent connections that use the same ports. Supporting these cross-layer dependencies through question-specific filtering is therefore an important direction for future work.

\section{Conclusions and future work}
\label{sec:conclusions}

This work introduces a reproducible methodology to evaluate LLMs on understanding network traffic idiosyncrasy. To that end, the  CICIDS2017 dataset is used and a set of specific questions, covering almost all the layers in the TCP/IP stack, was devised together with its human-driven generated ground truth answers. An LLM-based judge is in charge of automatically tag answers coming from the LLM under evaluation. Additionally, two data-preprocessing approaches were generated: default and protocol-filter view.  In general, the correctness remained limited and generally decreased with more packets, with Qwen~2.5~72B the most accurate model for both default and protocol-filter views. As expected, to be concise about the network traffic managed by the LLM, increases their understanding performance. Moreover, protocol filtering substantially reduces the prompt size thus reducing the number the number of tokens the LLM should manage and, consequently, the computation resources needed.

This study faces three main limitations: static, protocol-specific views; reliance on a single capture and annotator; and packet input sizes restricted by individual LLM context windows. To address them, future work should explore dynamic filter-selection by using agentic environments and broader evaluations featuring multiple captures and independent judges.

\section*{Acknowledgment}

This publication has been funded by MICIN/AEI/10.13039/501100011033 under grant PID2024-161902OB-I00.



\begin{thebibliography}{00}


\bibitem{Cui25}
T. Cui, X. Lin, S. Li, M. Chen, Q. Yin, Q. Li, K. Xu: “TrafficLLM: Enhancing Large Language Models for Network Traffic Analysis with Generic Traffic Representation,” arXiv:2504.04222, 2025 (DOI: \url{https://doi.org/10.48550/arXiv.2504.04222}).

\bibitem{Liu25}
F. Liu, B. Farkiani, P. Crowley: “Large Language Models for computer networking operations and management: A survey on applications, key techniques, and opportunities,” Computer Networks, vol. 271, article 111614, 2025 (DOI: \url{https://doi.org/10.1016/j.comnet.2025.111614}).

\bibitem{Fricke26}
S. Fricke, M. Friesen, J. Jasperneite: ”LLMs for network engineering automation: state of the art and research directions,” IEEE Network Operations and Management Symposium (NOMS), pp. 1-6, 2026 (DOI: \url{https://doi.org/10.1109/NOMS69089.2026.11668163}).

\bibitem{Nakamura26}
R. Nakamura, K. Eguchi: ”How helpful is LLM assistance in network operations? a case study at a large demonstration network,” IEEE Network Operations and Management Symposium, (NOMS), pp. 1-10, 2026 (DOI: \url{https://doi.org/10.1109/NOMS69089.2026.11668182}).

\bibitem{Paukkeri26}
S.-P. Paukkeri, T. Frantti: “Evaluation of local LLMs for network traffic packet analysis,” Journal of Cyber Security Technology, vol. 10, no. 1, 2026. (DOI: \url{https://doi.org/10.1080/23742917.2026.2671738}).

\bibitem{Ginige26}
Y. Ginige, B. Silva, T. Dahanayaka, S. Seneviratne: “TrafficLLM: LLMs for improved open-set encrypted traffic analysis,” Computer Networks, vol. 274, article 111847, 2026 (DOI:\url{https://doi.org/10.1016/j.comnet.2025.111847}).

\bibitem{Sharafaldin18}
I. Sharafaldin, A. H. Lashkari, A. A. Ghorbani: "Toward generating a new intrusion detection dataset and intrusion traffic characterization," Proceedings of the 4th International Conference on Information Systems Security and Privacy (ICISSP), pp. 108-116, 2018 (DOI: \url{https://doi.org/10.5220/0006639801080116}).

\bibitem{Soto25}
J. A. Delgado-Soto, J. E. López de Vergara, I. González, D. Perdices, L. de Pedro: “GPT on the wire: Towards realistic network traffic conversations generated with large language models,” Computer Networks, vol. 265, article 111308, 2025 (DOI: \url{ https://doi.org/10.1016/j.comnet.2025.111308}).

\bibitem{Enisoglu26}
R. Enisoglu, B. A. Inan: “Network traffic classification from handcrafted features to foundation models: A comprehensive survey,” Computer Networks, vol. 287, article 112578, 2026 (DOI: \url{https://doi.org/10.1016/j.comnet.2026.112578}).

\bibitem{Kholgh23}
D. K. Kholgh, P. Kostakos: "PAC-GPT: A novel approach to generating synthetic network traffic with GPT-3," IEEE Access, vol. 11, pp. 114936-114951, 2023 (DOI: \url{https://doi.org/10.1109/ACCESS.2023.3325727}).

\bibitem{Sharma23}
P. Sharma, V. Yegneswaran: "PROSPER: Extracting protocol specifications using large language models," Proc. 22nd ACM Workshop on Hot Topics in Networks (HotNets '23), pp. 41-47, 2023 (DOI: \url{https://doi.org/10.1145/3626111.3628205}).

\bibitem{Kan24}
K. B. Kan, H. Mun, G. Cao, Y. Lee: "Mobile-LLaMA: Instruction fine-tuning open-source LLM for network analysis in 5G networks," IEEE Network, vol. 38, no. 5, pp. 76-83, 2024 (DOI: \url{https://doi.org/10.1109/MNET.2024.3421306}).


\bibitem{Wang24}
D. Wu, X. Wang, Y. Qiao, Z. Wang, J. Jiang, S. Cui, F. Wang: "NetLLM: Adapting large language models for networking," Proceedings of the ACM SIGCOMM 2024 Conference, pp. 661-678, 2024 (DOI: \url{https://doi.org/10.1145/3651890.3672268}).

\bibitem{Kwon25}
D. Kwon, S. Kang, S. Choi: "TrafficNetQA: Question answering datasets for evaluating LLM performance on traffic network files," Proceedings of the 33rd ACM SIGSPATIAL International Conference on Advances in Geographic Information Systems, pp. 1142-1145, 2025 (DOI: \url{https://doi.org/10.1145/3748636.3763215}).

\bibitem{Maatouk26}
A. Maatouk, F. Ayed, N. Piovesan, A. De Domenico, M. Debbah, Z.-Q. Luo: "TeleQnA: A benchmark dataset to assess large language models telecommunications knowledge," IEEE Network, vol. 40, no. 2, pp. 253-260, 2026 (DOI: \url{https://doi.org/10.1109/MNET.2025.3576035}).

\bibitem{Marjou26}
X. Marjou, L. Tami, I. Jaffeux-Cheniout: "PCAP-LM: An LLM-native text representation for TLS bulk traffic analysis," arXiv:2607.28100, 2026 (DOI: \url{http://doi.org/10.48550/arXiv.2607.28100}).

\bibitem{Tothfalusi25} T. Tóthfalusi, Z. Csiszár and P. Varga, "TACO: A generative AI copilot for intent-based telecommunication core network analysis," in Proceedings of the 2025 IEEE Network Operations and Management Symposium (NOMS), 2025 (DOI: \url{http://doi.org/10.1109/NOMS57970.2025.11073654}).

\bibitem{Vico24}
R. Torre-Vico, R. Magán-Carrión, R.A. Rodríguez-Gómez: "Exploring the Use of LLMs to Understand Network Traces," 17th International Conference on Computational Intelligence in Security for Information Systems, pp. 122-131, 2024. Available at \url{https://link.springer.com/chapter/10.1007/978-3-031-75016-8_12}.

\bibitem{nesgRepo}
NESG Group, ``Do LLMs understand network traces? Code, prompts and results,'' GitHub repository. [Online]. Available: \url{https://github.com/NESG-group/llms-understand-network-traces}. Accessed: Nov. 9, 2026.

\bibitem{b14}
D. Li, B. Jiang, L. Huang, A. Beigi, C. Zhao, Z. Tan, A. Bhattacharjee, Y. Jiang, C. Chen, T. Wu, K. Shu, L. Cheng, H. Liu: ``From generation to judgment: opportunities and challenges of LLM-as-a-judge,'' Proceedings of the 2025 Conference on Empirical Methods in Natural Language Processing, pp. 2757--2791, 2025 (DOI: \url{https://doi.org/10.18653/v1/2025.emnlp-main.138}).



\bibitem{cohen1960}
J. Cohen: ``A coefficient of agreement for nominal scales,'' Educational and Psychological Measurement, vol. 20, pp. 37--46, 1960 (DOI: \url{https://doi.org/10.1177/001316446002000104}).

\bibitem{cohen1968}
J. Cohen: ``Weighted kappa: nominal scale agreement with provision for scaled disagreement or partial credit,'' Psychological Bulletin, vol. 70, no. 4, pp. 213--220, 1968 (DOI: \url{https://doi.org/10.1037/h0026256}).



\end{thebibliography}
\end{document}